\documentstyle[12pt]{article}

\begin{document}
\begin{titlepage}
\pagestyle{empty}
\topmargin-.5in
\baselineskip=14pt
\rightline{UMN-TH-1334/95}
\rightline{LBL-37019}
\rightline{UCB-95/109}
\rightline{hep-ph/yymmddd}
\rightline{April 1995}
\vskip .2in
\baselineskip=18pt
\begin{center}
{\large{\bf Preserving Flat Directions During Inflation}
}\footnote{This work was supported in part
by DoE grants
DE-FG02-94ER-40823 and DE-AC03-76SF00098, and by NSF grants AST-91-20005
and PHY-90-21139.}
\end{center}

\vskip .1in
\begin{center}

Mary K. Gaillard

{\it Department of Physics and Theoretical Physics Group, Lawrence Berkeley
Laboratory}

{\it University of California, Berkeley, California, 94720}

Hitoshi Murayama\footnote{On leave of absence from {\it Department of
Physics, Tohoku University, Sendai, 980 Japan}}

{\it Theoretical Physics Group, Lawrence Berkeley Laboratory}

{\it University of California, Berkeley, CA 94720, USA}

 and

Keith A. Olive

{\it School of Physics and Astronomy, University of Minnesota}

{\it Minneapolis, MN 55455, USA}

\vskip .1in

\end{center}
\vskip .2in
\centerline{ {\bf Abstract} }
\baselineskip=18pt
Supersymmetry is generally broken by the non-vanishing vacuum
energy density present during inflation. In supergravity models,
such a source of supersymmetry breaking typically makes a contribution
to scalar masses of order ${\tilde m}^2 \sim H^2$, where $H^2 \sim V/M_P^2$
is the Hubble parameter during inflation.  We show that in supergravity models
which possess a Heisenberg symmetry, supersymmetry breaking makes no
contribution
to scalar masses, leaving supersymmetric flat directions flat at
tree-level.  One-loop corrections in general lift the flat
directions, but naturally give small negative squared masses $\sim -
g^2 H^2/(4\pi)^2$ for all flat directions that do
not involve the stop.  No-scale
supergravity  of the SU(N,1) type and the untwisted sectors from orbifold
compactifications are special cases of this general set of
models.  We point out the importance of the preservation of flat directions
for baryogenesis.

\noindent
\end{titlepage}
\renewcommand{\thepage}{\roman{page}}
\setcounter{page}{2}
\mbox{ }

\vskip 1in

\begin{center}
{\bf Disclaimer}
\end{center}

\vskip .2in

\begin{scriptsize}
\begin{quotation}
This document was prepared as an account of work sponsored by the United
States Government. While this document is believed to contain correct
information, neither the United States Government nor any agency
thereof, nor The Regents of the University of California, nor any of their
employees, makes any warranty, express or implied, or assumes any legal
liability or responsibility for the accuracy, completeness, or usefulness
of any information, apparatus, product, or process disclosed, or represents
that its use would not infringe privately owned rights.  Reference herein
to any specific commercial products process, or service by its trade name,
trademark, manufacturer, or otherwise, does not necessarily constitute or
imply its endorsement, recommendation, or favoring by the United States
Government or any agency thereof, or The Regents of the University of
California.  The views and opinions of authors expressed herein do not
necessarily state or reflect those of the United States Government or any
agency thereof, or The Regents of the University of California.
\end{quotation}
\end{scriptsize}

\vskip 2in

\begin{center}
\begin{small}
{\it Lawrence Berkeley Laboratory is an equal opportunity employer.}
\end{small}
\end{center}

\newpage
\renewcommand{\thepage}{\arabic{page}}
\setcounter{page}{1}
\baselineskip=18pt

\def\la{~\mbox{\raisebox{-.6ex}{$\stackrel{<}{\sim}$}}~}
\def\ga{~\mbox{\raisebox{-.6ex}{$\stackrel{>}{\sim}$}}~}
\def\b-l{$B - L$}
\def\bl{$B + L$~}
\def\tm{$\tilde m$}
\def\beq{\begin{equation}}
\def\eeq{\end{equation}}
\topmargin0.0in
\setcounter{footnote}{0}
One of the more robust mechanisms for the generation of a cosmological
baryon asymmetry is the decay of the coherent scalar field oscillations
along flat directions in supersymmetric theories, commonly referred to as the
Affleck-Dine (AD) mechanism \cite{ad}.
It is well known that in the supersymmetric
standard model and in supersymmetric grand unified theories, there are
directions in field space in which both the $F$- and $D$-terms of the scalar
potential vanish identically. Such a direction was explicitly constructed
in the context of SU(5) in \cite{ad} and generalized to larger guts as well
\cite{m}.  Supersymmetry breaking spoils the flatness,
and one expects the generation of soft scalar masses for all scalar fields of
order of the supersymmetry breaking scale, $\tilde m$.
A baryon asymmetry can be generated if at some early stage in the evolution
of the Universe, the fields along the flat direction obtain large vacuum
expectation
values and can be associated with a baryon number violating operator.
Subsequently, when the expansion rate, given by the Hubble parameter, $H$, is
of order $\tilde m$, coherent scalar field oscillations along the flat
direction carrying
a baryon number begin. The decay of the scalar field oscillations results in a
final baryon asymmetry \cite{ad,lin}.

The AD scenario outlined above is normally implemented
in the context of inflation
\cite{eeno}.  During inflation, scalar fields with masses, $m < H$, are driven
by quantum fluctuations
to large vacuum values which become the source of the scalar field oscillations
once inflation is over.  To avoid the washout of the baryon asymmetry by
electroweak sphaleron interactions \cite{am}, the scenario works
most easily in the framework of a gut larger than minimal SU(5) in which $B-L$
is violated, though it is possible that the asymmetry can be preserved even
when $B-L$ is not violated \cite{dmo}. Indeed, it is possible to implement the
AD scenario for baryogenesis without any additional baryon number violation
beyond that
in the standard model so long as lepton number is violated \cite{cdo}.
Here, flat directions associated with lepton number violating operators can be
used to generate a lepton asymmetry which is subsequently converted into a
baryon asymmetry by sphaleron interactions.

Recently, it has been argued that the simple picture of driving scalar fields
to large vacuum values along flat directions during inflation is dramatically
altered
in the context of supergravity \cite{drt}.  During inflation, the Universe is
dominated by the vacuum energy density, $V \sim H^2 M_P^2$. The presence of a
non-vanishing and positive vacuum energy density indicates that supergravity is
broken and soft masses of order of the gravitino mass
$\sim H$ are generated \cite{bfs}.
A similar observation was made in \cite{dval} where it was noted that other
flat
directions, associated with moduli, also receive large contributions to their
masses from a non-zero vacuum energy density during inflation.  There it was
argued that in such cases there may be no moduli problem (a.k.a. Polonyi
problem \cite{pp}) involving excess entropy production.  However this solution
would require that the expectation value of the moduli are unchanged during
and after inflation, an unlikely circumstance.  The contribution to scalar
masses can be easily understood by considering
for example, the scalar potential in a supergravity model described
by a K\"ahler potential $G$ \cite{sugr},
\beq
V = e^G \left[ G_i {(G^{-1})}^i_j G^j - 3 \right]
\label{gen}
\eeq
where $G_i = \partial G/\partial \phi^i$ and $G^i =
\partial G/\partial \phi^*_i$.
A positive vacuum energy density, $V > 0$, needed for inflation,
requires that $e^G G_i \ne 0$, thus breaking supergravity, and implies
that the gravitino mass, $m_{3/2} = e^{G/2} \ne 0$.
Typically, scalar squared masses
will pick up a contribution of order $m_{3/2}^2 \sim V/M_P^2 \sim H^2$,
thus lifting the flat directions and potentially preventing the  realization of
the AD scenario as argued in \cite{drt}.  Below we use
reduced Planck units: $M_P/\sqrt{8\pi} = 1$.

To see this, let us consider a minimal supergravity model whose
K\"ahler potential is defined by
\beq
G = zz^* + \phi_i^* \phi^i + \ln |\overline{W}(z) + W(\phi)|^2
\label{min}
\eeq
where $z$ is a Polonyi-like field \cite{pol} needed to break supergravity, and
we denote the scalar components of the usual matter chiral
supermultiplets by $\phi^i$. $W$ and $\overline{W}$
are the superpotentials of $\phi^i$ and $z$ respectively. In this case,
the scalar potential becomes
\beq
V = e^{zz^* + \phi_i^* \phi^i} \left[ |\overline{W}_z +
z^* (\overline{W} + W)|^2
     + |W_{\phi^i} + \phi^*_i (\overline{W} + W)|^2 - 3|(\overline{W} + W)|^2
   \right]
\eeq
Included in the above expression for $V$, one finds a mass term for the matter
fields $\phi^i$, $e^G  \phi_i^* \phi^i = m_{3/2}^2 \phi_i^* \phi^i$.
 If $z$ breaks supergravity
this term is non-vanishing. As it applies to
all scalar fields (in the matter sector), all flat directions are lifted by
it as well.

Another instructive way to see the lifting of flat directions is to examine
the expression (see, for example \cite{tr}) for the trace of the scalar
squared mass matrix (we ignore $D$-terms assuming the inflaton is a
gauge singlet),
\beq
{\rm Tr} {\cal M}_0^2 = 2{\rm Tr} {\cal M}_{1/2}^2 + 2(N-1)V +
2m_{3/2}^2 \left[ (N-3) - G_i {(G^{-1})}^i_j R^j_k {(G^{-1})}^k_l G^l \right]
\label{trace} \eeq
where $(N-1)$ is the total number of chiral multiplets $\phi^i$,
$R_i^j = (\ln \det G)_i^j$ is the K\"ahler Ricci tensor, and
${\rm Tr} {\cal M}_{1/2}^2$ is the flat space value of the trace of the
chiral fermion squared mass matrix. In the minimal supergravity model
described by eq. (\ref{min}), $G_i^j = \delta_i^j$ and $R_i^j = 0$.
Thus the traces give
\beq
{\rm Tr}({\cal M}_0^2 - 2{\cal M}_{1/2}^2)_\phi + m_A^2 + m_B^2 - 2m_\chi^2 =
- 4m_{3/2}^2 + 2(N-1)\left(m_{3/2}^2 + V\right),
\label{smin}
\eeq
guaranteeing a splitting among each of the chiral multiplets $\phi$ by
an amount $m_{3/2}^2$ (neglecting space-time curvature contributions to
fermion masses).  In (\ref{smin}), $m_A$ and $m_B$ are the masses of the
scalar components of $z$, $\chi$ is its fermion component ({\it i.e.}, the
goldstino: $m_\chi = 2m_{3/2}$ at the ground state), and their masses
satisfy $m_A^2 + m_B^2 - 2m_\chi^2 =  - 4m_{3/2}^2$.
As shown below, a more careful definition of scalar ``masses'' is required when
we consider non-minimal supergravity models.

The above arguments can be generalized to supergravity models with
non-minimal K\"ahler potentials. By and large the same conclusions hold
as claimed in \cite{drt}. However as we will show below,
there is a wide class of models with considerable phenomenological interest
in which flat directions are not lifted by supergravity breaking.
Indeed, the flatness is ensured by a ``Heisenberg symmetry'' \cite{BG}
of the kinetic function in this class of models.

A special class of these models are no-scale supergravity models, that were
first introduced in \cite{ns} and have
the remarkable property that the gravitino mass is undetermined at the
tree level despite the fact that supergravity is broken.
No-scale supergravity has been
used heavily in constructing supergravity models in which all mass scales
below the Planck scale are determined radiatively \cite{nsguts},\cite{nshid}.
Indeed, the decoupling of the gravitino mass from
global supersymmetry breaking in the observable sector was used to
derive models with a gravitino mass at the Planck scale for the case
\cite{mpgrav} $g\equiv 0$ and within a few orders of that scale for the case
\cite{bg} $W_i\equiv 0$, with phenomenologically
acceptable supersymmetry breaking in the matter sector. A large gravitino mass
allows a sufficiently early decay of the gravitino, thus avoiding \cite{avoid}
potential cosmological
problems associated with the gravitino \cite{sw}. In this context, the moduli
problems can also be resolved \cite{enq}. Many inflationary models have also
been considered in the context of no-scale supergravity \cite{infl}.
These models emerge naturally in torus \cite{wit} or, for the untwisted sector,
orbifold \cite{orb} compactifications of the heterotic string.

We will consider the possibility that the inflaton is one of the $\phi^i$
or is a ``hidden sector'' field.  Thus we introduce the set of chiral fields
$z,\phi^i,y^a$, with $y^a$ in the hidden sector, and define the
``Heisenberg symmetry'' as follows~\cite{BG},
\begin{equation}
\delta z = \epsilon^*_i \phi^i, \hspace{1cm} \delta \phi^i = \epsilon^i ,
\hspace{1cm} \delta y^a = 0, \label{heis}
\end{equation}
where $\epsilon^i$ are complex parameters and $\epsilon_i^*$ their
complex conjugates.  The invariants under this symmetry are the $y^a$ and
the combination
\begin{equation}
\eta \equiv z + z^* - \phi_i^* \phi^i . \label{etadef}
\end{equation}
Let us assume this is a symmetry of the kinetic function in the K\"ahler
potential.  We also require that the field $z$ does not have a coupling
in the superpotential.
Then the most general K\"ahler potential becomes
\begin{equation}
G = f(\eta) + \ln |W(\phi)|^2 + g(y), \label{kpot}
\end{equation}
where the superpotential $W$ is a holomorphic function of $\phi^i$ only.

To analyze the resulting theory, we first look at the kinetic
terms of the chiral scalars.  In terms
of the original fields $z$ and $\phi^i$, the kinetic terms are written
as
\begin{equation}
K = f^{\prime\prime} |\partial z|^2
	+ f^{\prime\prime} \phi^i \partial z \partial \phi_i^*
	+ f^{\prime\prime} \phi^*_j \partial z^* \partial \phi^j
	+ (f^{\prime\prime} \phi^*_j \phi^i - f' \delta_j^i)
		\partial \phi^j \partial \phi_i^* .
\end{equation}
where $\prime$ denotes differentiation with respect to $\eta$.
To ``diagonalize'' the kinetic term so that chiral scalars do not
have mixing kinetic terms, we rewrite the kinetic terms in terms of
$\eta$ defined in Eq.~(\ref{etadef}) and a $U(1)$ current
\cite{Gelmini,bg,Linde}
\begin{equation}
I_\mu = i(\partial_\mu z - \partial_\mu z^*)
	-i (\phi^i \partial_\mu \phi_i^* - \phi_i^* \partial_\mu \phi^i) ,
\end{equation}
and they read as
\begin{equation}
K = f^{\prime\prime} \left[ (\partial \eta)^2 + (I_\mu)^2 \right]
	- f' |\partial \phi^i|^2 .
\label{kin}
\end{equation}
Therefore, we regard $\eta$ and $\phi^i$ as independent degrees of
freedom rather than $z$ and $\phi^i$.

We are now at the stage to write down the scalar potential.  Following
the general definition (\ref{gen}) and matrix inversion \cite{MSYY2},
\begin{equation}
V = e^{f(\eta)+g(y)} \left[
	\left( \frac{f^{\prime 2}}{f^{\prime\prime}} - 3 \right)
		|W|^2
	- \frac{1}{f'} |W_i|^2 + g_a(g^{-1})^a_bg^b\right ] .
\label{hpot}
\end{equation}
It is important to notice that the cross term $|\phi_i^* W|^2$ has
disappeared in the
scalar potential.  Because of the absence of the cross term, flat
directions remain flat even during the inflation.
A detailed discussion
of the dynamics of $\eta$ field during inflation can be found in
Ref.~\cite{MSYY2} for a specific choice of $f = (3/8) \ln \eta +
\eta^2$.  The no-scale model corresponds to $f = -3 \ln \eta$,
$f^{\prime 2} = 3 f^{\prime\prime}$ and the first term in (\ref{hpot})
vanishes. The potential then takes the form
\beq
V = e^g\left[e^{{2 \over 3}f} |W_i|^2 + e^fg_a(g^{-1})^a_bg^b\right],
\label{nspot}
\eeq
which is positive definite.  The requirement that the vacuum energy vanishes
implies \newline
$\langle W_i\rangle = \langle g_a\rangle $ at the minimum. As a
consequence $\eta$ is undetermined
and so is the gravitino mass $m_{3/2}(\eta)$.

The easiest way to see that the Lagrangian (\ref{hpot}) preserves the flat
direction is to look at the equation of motion of the fields $\phi^i$.  In the
inflationary universe, it reads as
\begin{equation}
\ddot{\phi}^i + 3 H \dot{\phi}^i + \Gamma^i_{j\eta}\dot\phi^j
\dot\eta	-{e^{f+g} \over f'}\left[
	\left( \frac{f^{\prime 2}}{f^{\prime\prime}} - 3 \right)
		W_i W^*
	- \frac{1}{f'} W_{ij}(W_j)^*\right] = 0,
\end{equation}
where the connection $\Gamma^i_{j\eta}$ is defined below.
Since the flat direction is characterized by $W_i = 0$, $W_{ij} = 0$, it
is easy to
see that a constant $\phi^i$ satisfies the equation of motion for any
values of $\phi^i$ along the flat direction.

It is probably also useful to look at the effective mass term for the
quantum fluctuations during inflation.  For definiteness, we
restrict ourselves to the no-scale case in the following discussion.
The usual
procedure of the covariant expansion around the background fields is
as follows.  We start from the general action
\begin{equation}
I[\phi] = \int d^4 x \sqrt{- g} \left[
	\frac{1}{2} g^{\mu\nu} G_{\alpha\beta}(\phi)
	\partial_\mu \phi^\alpha \partial_\nu \phi^\beta - V(\phi) \right],
\label{action} \end{equation}
and expand the fields as
\begin{equation}
\phi^\alpha = \phi^\alpha_0 + \xi^\alpha
	- \frac{1}{2} \Gamma^\alpha_{\beta\gamma} \xi^\beta \xi^\gamma
	+ {\cal O}(\xi^3).
\end{equation}
Here and below, the indices $\alpha, \beta, \ldots$ refer generically to
$\eta$, $\phi^i,\;\phi^*_i$, and the Christoffel symbol
$\Gamma^\alpha_{\beta\gamma}$ is derived from the metric
$G_{\alpha\beta}(\phi)$ on the field space.  We obtain
\begin{eqnarray}
\lefteqn{ I[\phi] = I[\phi_0]
	+ \int d^4 x \sqrt{- g} \left[ g^{\mu\nu} G_{\alpha\beta}
		\partial_\mu \phi_0^\alpha D_\nu \xi^\beta
		- \partial_\alpha V \xi^\alpha \right] } \nonumber \\
	& & + \frac{1}{2} \int d^4 x \sqrt{- g} \left[
		g^{\mu\nu} G_{\alpha\beta}
		D_\mu \xi^\alpha D_\nu \xi^\beta
		+ R_{\alpha\gamma\delta\beta} \xi^\gamma \xi^\delta
			g^{\mu\nu} \partial_\mu \phi_0^\alpha
			\partial_\nu \phi_0^\beta
		- (D_\alpha \partial_\beta V) \xi^\alpha \xi^\beta
		\right] \nonumber \\
		& & + {\cal O} (\xi^3) .
\end{eqnarray}
The covariant derivative is defined by $D_\alpha \xi^\beta =
\partial_\alpha \xi^\beta + \Gamma^\beta_{\alpha\gamma} \xi^\gamma$.
The last term contains the effective term mass term for the fluctuation
$\xi^\alpha$,
\begin{equation}
(m^2)_{\alpha\beta} = D_\alpha \partial_\beta V
	= \partial_\alpha \partial_\beta V
	- \Gamma^\kappa_{\alpha\beta} \partial_\kappa V ,
\end{equation}
and its trace gives the field reparametrization invariant result (\ref{trace}),
with $R^j_k = {1\over3}(N+1)G^j_k$ in the no-scale case.
We now look at the fluctuation $\xi^i$ of the $\phi^i$
fields.  In our case, however, we assume that the $\eta$ field is
constant during inflation and do not use the equation of motion for
the $\eta$ field.  If we used the equation of motion, the $\eta$ field
would roll down the potential, and the potential identically vanishes
in the limit $\eta \rightarrow \infty$.  It is usually assumed that
higher order corrections to the potential stops this run-away
behavior, and $\eta$ will be fixed at some point. How this may explicitly be
realized will be discussed below. Following this standard approach and
regarding
the $\eta$ field as a constant, we do not allow a fluctuation for $\eta$.
This in turn means $\xi^\eta - \Gamma^\eta_{ij} \xi^i \xi^j/2 = 0 + O(\xi^3).$
Then the linear term in the $\xi^\eta$ does not vanish in $I[\phi]$,
but precisely cancels the second term $\Gamma^\eta_{ij}
\partial_\eta V$ in $(m^2)_{ij}$.  The effective mass term is hence
just $\partial_i \partial_j V$.  Then it should be clear from the
explicit  form of the potential $V$ that there is no additional mass
term to the flat direction even during inflation.\footnote{For a
more general choice of $f(\eta)$ as in \cite{MSYY2}, the potential may
not have this
run-away behavior for $\eta$.  In this case, $\eta$ will
settle to its minimum during inflation, and $V_\eta$ vanishes there.
Therefore, the effective mass term is also just $\partial_i \partial_j
V$ in the more general case as well.}

A natural question arises as to whether higher order corrections modify the
form of the K\"ahler potential, thereby lifting the flat direction.
Indeed, only the kinetic energy part of our K\"ahler potential respects
the Heisenberg symmetry, which is broken by gauge couplings as well as by the
superpotential.  However, an explicit one-loop
calculation shows that the gravitational interactions preserve the
Heisenberg symmetry \cite{BG}.  Therefore, the only possible
contribution to the mass of the flat directions come from either gauge
interactions or superpotential couplings that contribute to the renormalized
K\"ahler potential.  Using the general results~\cite{tr,TR} for the
one-loop corrected supergravity lagrangian, we obtain,
assuming that inflation is driven by an $F$-term
rather than a $D$-term (the result is similar if $\langle D\rangle\ne 0$):
\begin{eqnarray}
\left( m^2 \right)_i^j &=& \frac{\ln(\Lambda^2/\mu^2)}{32\pi^2}\Bigg\{ h_{ikl}
h^{*jkl} \left[\alpha\langle V \rangle + m_{3/2}^2\left(5{f'^2\over f''} +
2{f'f'''\over f''^2} - 10 - {f'^4\over f''^2}\right)
 \right] \nonumber \\ & &
- 4\delta^j_ig_a^2C_2^a(R_i)\left[\beta\langle V \rangle
+ m_{3/2}^2\left({f'f'''\over f''^2} - 2\right) \right]\Bigg\} ,
\label{mass}\end{eqnarray}
where the vacuum energy $\langle V \rangle $ and the gravitino mass $m_{3/2}$
are their values during inflation. $\mu\ge \langle V\rangle $ is the
appropriate infrared cut-off in the loop integral, and
$\Lambda$ is the cutoff scale below which the effective supergravity
Lagrangian given by the K\"ahler potential eq.~(8) is valid; $\Lambda =
1$ in many models but can be lower if the inflaton is a composite field.
The $h$'s are Yukawa coupling constants
$g_a,C_2^a(R_i)$ are the coupling constant and matter Casimir for the
factor gauge group $G_a$, and the parameter $\alpha,\beta,$ are model
dependent.  In the no-scale case $f = - 3\ln\eta$, the result
(\ref{mass}) reduces to\footnote{In the context of string theory, Eq.
(\ref{nsmass}) is valid
in more realistic multi-moduli no-scale models that describe the untwisted
sectors from orbifold compactifications.}
\begin{equation}
\left( m^2 \right)_i^j = \frac{\ln(\Lambda^2/\mu^2)}{32\pi^2}\langle V \rangle
\left[\alpha h_{ikl} h^{*jkl} - 4\beta\delta^j_ig_a^2C_2^a(R_i) \right].
\label{nsmass}\end{equation}
If the inflaton is one of the $\phi^i$ ($\phi_0\ne 0$), $\alpha = \beta =
{2\over3}$. If the
inflaton is in the hidden sector ($y_a\ne 0$), $\alpha < 0,\;
\beta = 1$ if the inflaton is not the dilaton $s: \;\langle s\rangle =
g^{-2}$.
If the inflaton is the standard string dilaton, $\beta = \alpha = -1$.
In all but the last case the masses are negative if gauge couplings dominate
Yukawa couplings.  In the following we assume the inflaton is not the dilaton.

There are several interesting points in the above formulae.  First of all,
higher order effects are always suppressed by the typical one-loop factor
$1/(4\pi)^2$.  The vacuum energy is related to the expansion rate by
$H^2 = (1/3) \langle V \rangle$, and hence the typical mass of the flat
directions is $m^2 \simeq 10^{-2} H^2$ during inflation. While these
one-loop masses are small enough to generate classical fields on large scales
during inflation, they are too large to allow sufficient growth due
to quantum fluctuations
in order to generate a sizeable baryon asymmetry.
During inflation, quantum fluctuations in $\langle \phi^2 \rangle$
begin to grow in time as $H^3 t /4 \pi^2$ \cite{fluct} up to a limiting
value given by $\langle \phi^2 \rangle = 3H^4/8\pi^2m^2$. During inflation,
the low momentum modes of these fluctuations will be indistinguishable
from a classical field with an amplitude $\phi_0 \simeq
\sqrt{\langle \phi^2 \rangle}$. In the (AD) mechanism for baryogenesis with
inflation, the baryon asymmetry produced is \cite{eeno}
\beq
\frac{n_B}{s} \sim  \frac{\epsilon {\phi_0}^4 {m_I}^{3/2}}
{{M_X}^2 {M_P}^{5/2} \tilde{m}}
  \la 10^{12} ({\phi_0 \over M_P})^4
\eeq
where $m_I \sim 10^{-7} M_P$ is the inflaton mass, $M_X \sim 10^{-3} M_P$
is the scale associated with the baryon number violation,
$\tilde{m} \sim 10^{-16}$ is the susy breaking scale
when $V=0$, i.e. after inflation, and $\epsilon$
is a measure of the CP violation and in this case can easily be O(1).
If ${\phi_0}^2$ were given by the maximum value of the fluctuations
generated by inflation, then ${\phi_0}^2 \sim 6 H^2 \sim 10^{-13}M_P^2$
and too small an asymmetry would result.
However, as we have seen in eq. (\ref{nsmass}), for all scalar
matter fields aside from stops, the contribution
to the mass squared is {\em negative} as the Yukawa couplings are smaller than
the gauge couplings.  Thus any flat direction not involving stops, will have a
negative contribution at one-loop without an ad hoc choice of the
parameters.\footnote{We do not need to choose a particular set of
parameters because the flat direction is preserved at tree-level,
and the one-loop correction is a small perturbation.
In models where the flat directions are lifted at tree level, one has to add
by hand tree-level terms
like $\delta G = c |\phi|^2 |\psi|^2$ to the K\"ahler potential,
and impose~\cite{drt} $c < 0$, where $\psi$ is
the inflaton, in order to generate a total negative squared mass
(in minimal supergravity, the tree level squared masses that are
generated during inflation are positive as seen in eq.(\ref{smin})).  However,
this is an {\cal O}(1) effect, and one has to be
careful not to spoil the positivity of the potential and the kinetic
terms.  For instance, one has to impose $-1 < c < -2/3$ for the case of
the inflaton potential in \cite{Holman}.}
Now, even though fluctuations
will begin the growth of $\phi_0$, the classical equations of motion
soon take over.  The classical equations of motion drive $\phi_0$ as
$(- m^2) t$ which is smaller than the quantum growth only for
$Ht < H^4/m^4$.  Then for $Ht > H^2/(- m^2)$, the classical growth of
$\phi_0$, becomes nonlinear $\sim H e^{-m^2 t/3H}$,
and $\phi_0$ will run off to its minimum
determined by the one-loop corrections to $\phi^4$, which are again
of order $V$.  An explicit one-loop calculation~\cite{tr,TR} shows that the
effective potential along the flat direction has a form
\begin{equation}
V_{eff} \sim \frac{g^2}{(4\pi)^2}
\langle V \rangle \left(
	-2 \phi^2 \log \left(\frac{\Lambda^2}{g^2 \phi^2}\right)
	+ \phi^2 \right)
	+ {\cal O}(\langle V \rangle)^2 ,
\end{equation}
where $\Lambda$ is the cutoff of the effective supergravity theory, and
has a minimum around $\phi \simeq 0.5 \Lambda$.  Also this is
consistent because this effective
potential is only of order $- \langle V \rangle g^2/(4\pi)^2$ and is a
small correction to the inflaton energy density which drives the inflation.
Thus, $\phi_0 \sim M_P$ will be generated and
in this case the subsequent sfermion oscillation will
dominate the energy density and a baryon asymmetry will result
which is independent of inflationary parameters as originally discussed in
\cite{ad,lin} and will produce $n_B/s \sim O(1)$.

Finally, as noted above, in order to realize the scenario presented here in the
no-scale case, we need a mechanism to stabilize the $\eta$-field.
In no-scale models, typically $\langle
V\rangle=0,\;m^2_{3/2}\ne 0$ at the
ground state.  Therefore it is very plausible that there is a solution with
$\langle V\rangle\sim \ln(\Lambda^2/m^2_{3/2})m^4_{3/2}/32\pi^2\ll m^2_{3/2}$
during inflation.  Under this assumption,
the one loop corrections \cite{tr} give a contribution
\beq
\Delta V = -{1\over 32\pi^2}\left[\alpha m_{3/2}^4\ln(\Lambda^2/m_{3/2}^2)
+ O(V^2)\right], \eeq
with $m^2_{3/2} \propto e^{f(\eta)}$, and $\alpha$ is model-dependent. If
$\phi_0$ is the inflaton, $\alpha = 14$, for a hidden sector inflaton $y_o$,
$\alpha = 13$--$21$.\footnote{If there are $N_h$ additional hidden sector
scalars with masses of order $m_{3/2}$, $\alpha \to \alpha + N_h$; here we set
$N_h = 0.$}  For a hidden sector inflaton, $V \propto e^{f(\eta)}$ and
the requirements that $V + \Delta V > 0, \; m^2_\eta > 0$, give
give the condition $3/2 > \ln(\Lambda^2/m_{3/2}^2) > 1$ which could be
considered as a fine tuning condition.  Actually,
these values correspond to results found (see, {\it e.g.}~\cite{nshid,bg})
in models where the
symmetry breaking potential is generated by condensation at a scale
$\Lambda$. For an inflaton $\phi_0, \; V \propto e^{f(\eta)/3}$, and the
$m_\eta^2$ mass is always positive; positivity of the one-loop corrected
potential requires only $\ln(\Lambda^2/m_{3/2}^2) > 5/3$  Once the positivity
requirements are satisfied, we obtain $m^2_\eta \sim \langle V\rangle$,
which is sufficient  to assure that inflation can occur in the false vacuum.

    In summary, we have shown that although it is a relatively
general property that flat directions receive tree-level masses during
inflation, this conclusion does not apply to the a class of models
which possess a Heisenberg symmetry. This class contains the
 phenomenologically interesting no-scale supergravity models
as well the forms of supergravity expected from string theory
truncations. We have also shown that although flat directions
remain flat at the tree-level in this class of models, one-loop
corrections upset the flatness and we expect that for
 all flat directions which do not involve stops, a negative
 mass squared is generated reulting in large expectation values
along the flat directions, leading, in turn, to (AD) baryogenesis along the
 lines originally suggested.

\vskip 0.8truecm
\noindent {\bf Acknowledgements}
\vskip 0.4truecm
We would like to thank Bruce Campbell for helpful conversations.
This work was supported in part by the Director, Office of Energy
Research, Office of High Energy and Nuclear Physics, Division of High
Energy Physics of the U.S. Department of Energy under Contracts
DE-FG02-94ER-40823 and
DE-AC03-76SF00098, by NSF grants AST-91-20005 and PHY-90-21139.

\end{document}